\documentclass[journal,transmag]{IEEEtran}

\usepackage{cite}
\usepackage{xfrac}
\usepackage{amsmath,esint}
\usepackage{amsfonts} \usepackage{amsthm}
\usepackage{caption}
\usepackage{subcaption}
\usepackage{color}
\usepackage{mathtools}
\usepackage{accents}
\newcommand*\VF[1]{\mathbf{#1}}
\newcommand*\dif{\mathop{}\!\mathrm{d}}

\newcommand*\iunit{\mathtt{i}}
\usepackage{marvosym}
\usepackage{amsthm}

\usepackage{hyperref} \usepackage{url}
\usepackage{listings} 

\usepackage[table,xcdraw]{xcolor}
\usepackage[normalem]{ulem}
\useunder{\uline}{\ul}{}
\usepackage{graphicx}
\graphicspath{ {./figures/} }
\usepackage{caption}
\usepackage{subcaption}
\usepackage{array}
\newcommand{\normal}{\mathbf{n}}
\newcommand{\bbX}{\mathbb{X}}
\newcommand{\curl}{\VF{curl}}
\newcommand{\divergence}{\operatorname{div}}

\makeatletter
\newcommand\suchthat{\@ifstar
  {\mathrel{}\middle|\mathrel{}}
  {\mid}}
\makeatother

\begin{document}

\title{Finite Element Analysis of Generalized Symmetric Formulations of Linear Maxwell's Equations in 3D Bianisotropic Media Permitting Magnetic Monopole Charges}
\title{Finite Element Methods for Linear Maxwell's Equations in Bianisotropic Media Permitting Polarization Fields and Magnetic Currents}

\author{\IEEEauthorblockN{Tharindu Fernando\IEEEauthorrefmark{1},
Martin Licht\IEEEauthorrefmark{2}, and
Michael Holst\IEEEauthorrefmark{3}}
\IEEEauthorblockA{\IEEEauthorrefmark{1}Department of Physics, University of Washington, Seattle, WA 98195 USA}
\IEEEauthorblockA{\IEEEauthorrefmark{2}Department of Mathematics, EPFL, 1015 Lausanne, Switzerland}
\IEEEauthorblockA{\IEEEauthorrefmark{3}Department of Mathematics, University of California, San Diego, CA 92093 USA}\thanks{Corresponding author: Tharindu Fernando (email: tharindu@uw.edu).}
}

\markboth{Journal of \LaTeX\ Class Files,~Vol.~14, No.~8, August~2015}{Shell \MakeLowercase{\textit{et al.}}: Bare Demo of IEEEtran.cls for IEEE Transactions on Magnetics Journals}

\IEEEtitleabstractindextext{\begin{abstract}
We review Maxwell's equations and constitutive relations for 3D bianisotropic media in a generalized form:
we consider all four variables and allow for nonzero polarization or magnetization, and also nonzero nonzero magnetic charge or current.
After a discussion of general boundary conditions, 
we introduce a time-harmonic variational formulation of linear Maxwell's equations within 3D bianisotropic media
in terms of the electric and magnetic fields.
We showcase a finite element approximation of our variational formulation,
using curl-conforming N\'ed\'elec edge elements of the first kind.
Numerical examples illustrate the convergence of the method.
\end{abstract}
}

\maketitle

\IEEEdisplaynontitleabstractindextext

\IEEEpeerreviewmaketitle

\newpage

\section{\label{sec:intro} Introduction}
\IEEEPARstart{M}{axwell's} equations are the constitutional system of partial differential equations for classical electromagnetism. They describe how electric and magnetic fields (components of a single electromagnetic field) evolve in terms of charges, currents, and material parameters. 
In this contribution, we approach the theoretical and numerical analysis of a Maxwell system that allows for nonzero magnetic charges and currents, nonzero polarization and magnetization, and bianisotropic media. We henceforth refer to this setup as a \emph{generalized} Maxwell system.
We demonstrate that contemporary numerical techniques are capable of solving such generalized electromagnetic problems.
This is in contrast to the literature on Maxwell's equations, which tends to assume several simplifications, focusing on simplified Maxwell systems in the vacuum setting in which analytical techniques provide explicit solutions. To our understanding, the numerical literature tends to adopt this simplified setup despite having the capability to explore more complicated problems.

We focus on a two-variable formulation of Maxwell's equations involving the electric field $\VF{E}$
and the magnetic field $\VF{H}$, which has seen increased interest in recent research
\cite{anees2018mixed,anees2018time,anees2019time,angermann2019semi,daveau1999new,mackie1993three,zyserman2000parallel}. 
We describe a mixed finite element method that discretizes both variables with N\'ed\'elec edge elements of the first kind. While rigorous \emph{a~priori} error analysis for this method is beyond the scope of this work, our numerical experiments indicate that this method converges as the mesh size decreases.

The idea of nonzero magnetic charges and currents has appealing potential applications.
For example, recent research has shed more light on magnetic monopoles using spin ice systems~\cite{castelnovo2008magnetic,bramwell2009measurement,morris2009dirac}.
Although in this case Maxwell's equations still take their standard form with no magnetic charges, 
different models with magnetic charges may be studied to influence new experimental setups
or to understand the implications if magnetic charges were observed in nature. 

For example, condensed matter physics uses analogue Maxwell systems that include Dirac monopoles~\cite{gangaraj2017berry}. 
Studying these Dirac monopoles is of paramount importance in quantum physics 
because of their physical realizations in state-of-the-art technologies involving semiconductors.
Unlike the real-space fields governed by Maxwell's equations,
Dirac monopoles occur in the so-called momentum space, which relates to real-space via the Fourier transformation.

The remainder of this work is structured as follows.
We introduce Maxwell's equations in their original four-variable form in Section~\ref{sec:maxwell}.
We discuss constitutive relations in Section~\ref{sec:constitutive}
and boundary conditions in Section~\ref{sec:bc}.
We present our variational theory in Section~\ref{sec:variational}. 
Finally, we discuss the results of numerical computations in Section~\ref{sec:numerics}.

\section{\label{sec:maxwell} Maxwell's equations}
The four equations of the generalized Maxwell system in \emph{phasor} or \emph{time-harmonic} form read as:
\begin{align} 
 \label{eq:gauss}
 \nabla\cdot{\VF{D}} &={\rho}_{E}, \\
 \label{eq:magnetic_gauss}
 \nabla\cdot{\VF{B}} &={\rho}_{M}, \\
 \label{eq:ampere}
 \nabla\times{\VF{H}} &= \iunit \omega{\VF{D}}  + {\VF{J}}_{E}, \\
 \label{eq:faraday} 
 - \nabla\times{\VF{E}} &= \iunit \omega{\VF{B}} + {\VF{J}}_{M}.
\end{align} 
Here, $\iunit$ denotes the imaginary unit and $\omega$ denotes the angular frequency. 
Equation~\eqref{eq:gauss} is the \emph{electric Gauss's law} 
and is commonly called \emph{Gauss's law}.
It describes the relationship 
between the \emph{electric flux density} $\VF{D}$,
and the \emph{eletric charge density} $\rho_{E}$.  
The \emph{magnetic Gauss's law}~\eqref{eq:magnetic_gauss} concerns 
the \emph{magnetic flux density} $\VF{B}$,
and the \emph{magnetic charge density} $\rho_{M}$.
The \emph{Amp\`ere-Maxwell law}~\eqref{eq:ampere}
relates three quantities:
the electric flux density $\VF{D}$,
the \emph{magnetic field} $\VF{H}$
and the \emph{electric current density} $\VF{J}_{E}$. 
\emph{Faraday's law}~\eqref{eq:faraday} relates 
the magnetic flux density $\VF{B}$,
the \emph{electric field} $\VF{E}$,
and the \emph{magnetic current density} $\VF{J}_{M}$. 

Even though $\rho_{M} = 0$ and $\VF{J}_{M} = 0$ in our current
understanding of physics, we consider nonzero $\rho_{M}$ and $\VF{J}_{M}$
for the purpose of mathematical inquiry.

The time-harmonic formulation treats the quantities 
$X \in (\VF{E}, \VF{H}, \VF{D}, \VF{B}, \VF{J}_E, \VF{J}_M, \rho_E, \rho_M)$
as sinusoidal as an assumption \emph{a priori}.
Then the real-time form $\hat{X}(\VF{r},t)$ of the field ${X}(\VF{r},\omega)$ in phasor form 
satisfies 
\begin{equation} \label{eq:rem:phasor:1}
    \hat{X}(\VF{r},t) = \operatorname{Re}[X(\VF{r},\omega) e^{\iunit \omega t}],
\end{equation}
where $\VF{r}$ are the spatial coordinates and $t$ denotes time.
The phasor formulation conveniently replaces the time derivatives of all fields $X$ by $\iunit \omega$.
In this work, we assume all $X$ to be in phasor form unless otherwise specified. 
\\

Taking the divergence of Amp\`ere's law~\eqref{eq:ampere} and Faraday's law~\eqref{eq:faraday},
and using the two Gauss's laws~\eqref{eq:gauss}-\eqref{eq:magnetic_gauss} 
yields two more identities,
known as the \emph{electric continuity equation}
and the \emph{magnetic continuity equation}, respectively:
\begin{gather}
 \label{eq:continuityequation:electric:cnty}
 \iunit \omega\rho_{E} + \nabla \cdot \VF{J}_{E} = 0,
 \\
 \label{eq:continuityequation:magnetic:cnty}
 \iunit \omega\rho_{M} + \nabla \cdot \VF{J}_{M} = 0.
\end{gather}
They describe the \emph{conservation of charge},
as the charge does not change over time in the absence of an electric or magnetic current.
The continuity equations are necessary conditions on the data for the existence of solutions to a Maxwell system.

\section{\label{sec:constitutive} Constitutive relations}
When an electromagnetic field is applied to a material, the bound charges and currents of the material respond.
This response is defined by \emph{constitutive relations}. 
These relations effectively describe how the contributions of the electric and magnetic fields in propagating electromagnetic waves change within different materials.
We consider the following general constitutive relations: 
\begin{align}
\label{eq:genconst0a} \VF{D}&=\epsilon\VF{E}+ \xi\VF{H} + \VF{P},\\
\label{eq:genconst0b} \VF{B} &= \zeta\VF{E} + \mu\VF{H} + \VF{M}.
\end{align}
Here, we use the \emph{electric polarization field} (or simply \emph{polarization}) $\VF{P}$ 
and the \emph{magnetic polarization field} (or \emph{magnetization}) $\VF{M}$.
The terms $\epsilon$, $\xi$, $\zeta$, and $\mu$ denote coefficient tensors that depend on the material.

In this work, we treat $\VF{P}$ and $\VF{M}$ as data.
In general, however, they may depend on the field variables $\VF{E}$ and $\VF{H}$ and thus be sources of nonlinearity.
For instance, when we have a Taylor expansion $\VF{P} = \sum_{n=1}^{\infty} A_n : \otimes^{n} \VF{E}$
of the electric polarization in terms of the electric field, 
then truncating up to the first-order term describes \emph{Pockel's effect},
and truncating up to the second-order term describes the \emph{Kerr effect}.
Although such effects are interesting in their own right,
we freeze $\VF{P}$ and $\VF{M}$ in this work to study the general linear problem.
We remark that solving linear generalized Maxwell's equations is not only an important theoretical stepstone towards developing nonlinear models in electromagnetism:
\emph{linearized} nonlinear Maxwell's equations appear frequently as auxiliary computations in numerical algorithms.

Without further assumptions on the material,
the coefficients $\epsilon$, $\xi$, $\zeta$, and $\mu$ are tensorial 
(thereby depending on the direction of the material), 
and we say that the material is \emph{bianisotropic}.
If the material coefficients are not tensorial, the material is called \emph{biisotropic}.
We refer the reader to 
\cite{lukas2013ETHnotes,jackson1999classical,dmitriev2000constitutive}
for more details on constitutive relations and bianisotropic media.
Although we will present numerical results only for the biisotropic case,
an extension to bianisotropic media is straightforward
when considering a separate Maxwell equation for each nonzero tensor component, for instance.
\\
    
As an example of this approach, we conceptualize the vacuum as a linear, homogeneous, and isotropic medium.
Here, the constitutive relations simplify to: 
\begin{align*}
    \VF{D} = \epsilon_{0}\VF{E}, \quad \VF{B} = \mu_{0}\VF{H}.
\end{align*} 
The values of $\epsilon_{0}$ and $\mu_{0}$ are known to be $\epsilon_{0}\approx 8.85\times 10^{-12} [\sfrac{Farad}{meter}]$ (the \emph{permittivity of free space})
and $\mu_{0}= 4\pi\times 10^{-7} [\sfrac{Newton}{Ampere^2}]$ (the \emph{permeability of free space}). 
Although this formulation occurs frequently in the literature, we use the bi-isotropic form in this work for broader generality.

\section{\label{sec:bc} Boundary conditions}
Since Maxwell's equations are partial differential equations, 
one requires boundary conditions to arrive at solutions. 
\emph{Interface conditions}, also known as \emph{general boundary conditions}, describe electromagnetic fields at the intersection of two materials. In this work, we consider the common \emph{perfectly conducting boundary conditions}~\cite{angermann2019semi},
where the electromagnetic fields inside one material are zero.

Let $\Omega$ be the region of integration with boundary $\partial\Omega$, 
and let $\normal$ be the unit normal pointing from the first material to the second.
Recall that the dot product of $\normal$ with a vector field along $\partial\Omega$ gives the magnitude of the field's normal component,
whereas the cross product between the two gives the field's tangential component. 
We assume that the boundary $\partial\Omega$ is split into two complementary parts $\Sigma^{E}$ and $\Sigma^{M}$.
The boundary conditions read 
\begin{align} 
\label{eq:bvp:EH} 
\begin{split}
\normal\times\VF{E}&=\VF{K_M}      \textrm{ along } \Sigma^{E},
	\\
\normal\times\VF{H}&=\VF{K_E}      \textrm{ along } \Sigma^{M},
	\\
\normal\cdot\VF{D}&=\VF{\sigma_E}  \textrm{ along } \Sigma^{M},
	\\
\normal\cdot\VF{B}&=\VF{\sigma_M}  \textrm{ along } \Sigma^{E}.
\end{split}
\end{align} 
Here, 
$\VF{K_E}$ is the \emph{surface electric current density},
$\VF{\sigma_E}$ is the \emph{surface electric charge density},
$\VF{K_M}$ is the \emph{surface magnetic current density},
and
$\VF{\sigma_M}$ is the \emph{surface magnetic charge density}.

We remark that although $\VF{K_M}=0=\VF{\sigma_M}$ in nature~\cite{jackson1999classical,griffiths2005introduction},
we consider nonzero magnetic currents and charges for the sake of generality, using the symmetric form given by~\eqref{eq:bvp:EH}.

\section{\label{sec:variational} Variational theory}
We derive a version of Maxwell's equations that is solely in terms of $\VF{E}$ and $\VF{H}$.
This is achieved by substituting the constitutive relations~\eqref{eq:genconst0a}-\eqref{eq:genconst0b}
into the Maxwell relations~\eqref{eq:gauss}-\eqref{eq:faraday}.
This gives us the differential form of Maxwell's equations in terms of only $\VF{E}$ and $\VF{H}$, which we call the \emph{($\VF{E}$, $\VF{H}$) Maxwell system}.
It comprises the Amp\`ere-Maxwell law and Faraday's laws:
\begin{align} 
 \label{eq:ampereEH} 
 \left[ \iunit\omega \left( \epsilon \VF{E} \right) + \iunit\omega \left( \xi \VF{H} \right) \right] - \nabla\times\VF{H} &= -\VF{J}_{E} - \iunit\omega \VF{P},
 \\
 \label{eq:faradayEH} 
 \left[ \iunit\omega \left( \zeta \VF{E} \right) + \iunit\omega \left( \mu \VF{H} \right) \right] + \nabla\times\VF{E} &= - \VF{J}_{M} - \iunit\omega \VF{M}.
\end{align} 
Taking the divergence of~\eqref{eq:ampereEH} and~\eqref{eq:faradayEH} 
together with the continuity equations~\eqref{eq:continuityequation:electric:cnty} and~\eqref{eq:continuityequation:magnetic:cnty}
produces the compatibility conditions 
\begin{align} 
 \label{eq:electricgaussEH}
 \nabla \cdot \left(  \epsilon \VF{E} \right) + \nabla \cdot  \left( \xi \VF{H} \right) &= \rho_{E} - \nabla \cdot \VF{P},
 \\
 \label{eq:magneticgaussEH}
 \nabla \cdot \left(  \zeta \VF{E} \right) + \nabla \cdot \left( \mu \VF{H} \right) &= \rho_{M} - \nabla \cdot \VF{M}. 
\end{align} 
We hence focus on the Amp\`ere-Maxwell law and Faraday's law.
\\

For the discussion of a variational formulation that is amenable to finite elements and their error analysis, 
we introduce a few notions of function spaces. 
For any domain $\Omega \subseteq \mathbb{R}^3$,
we write $L^{p}(\Omega)$ for the Lebesgue space to exponent $1 \leq p \leq \infty$,
and $\|\cdot\|_{p,\Omega}$ denotes the associated norm. 
The case $p=2$ is the most important case: $L^{2}(\Omega)$ is the Hilbert space of square-integrable functions.

We recall the Sobolev spaces~(see~\cite{stakgold2011green,monk2019finite,PeterMonk2003})
\begin{align} \label{eqs:sobolevspaces}
    H(\curl;\Omega) &= \{ \VF{v} \in L^2(\Omega)^{3} \suchthat \nabla\times \textrm{ }\VF{v} \in L^2(\Omega)^{3} \}
    ,
    \\
    H(\divergence;\Omega)  &= \{ \VF{v} \in L^2(\Omega)^{3} \suchthat \nabla\cdot \textrm{ }\VF{v} \in L^2(\Omega)^{3} \}
    .
\end{align} 
These are equipped with the respective norms 
\begin{gather*} 
    \| \VF{X} \|_{\curl,\Omega} := \| \VF{X} \|_{2,\Omega} + \| \nabla \times \VF{X} \|_{2,\Omega},
    \\
    \| \VF{X} \|_{\divergence,\Omega} := \| \VF{X} \|_{2,\Omega} + \| \nabla \cdot \VF{X} \|_{2,\Omega}.
\end{gather*}
For a rigorous discussion of boundary conditions, we recall that these spaces have well-defined tangential and normal traces
on open subsets of the boundary $\partial\Omega$.
We write $\langle\cdot,\cdot\rangle_{\Omega}$ for the $L^{2}$ product of vector fields over $\Omega$.
Conceptually, 
when $\VF{X}$ and $\VF{\phi}$ are sufficiently smooth vector fields and the geometry is sufficiently regular,
we recall the integration by parts formula
\cite{jackson1999classical,monk2019finite}: 
\begin{equation} \label{eq:result:ibp:curl}
    \langle \nabla\times\VF{X}, \VF{\phi} \rangle_{\Omega}
    =
    \langle \VF{X}, \nabla\times\VF{\phi} \rangle_{\Omega}
    +
    \langle \normal\times\VF{X}, \VF{\phi} \rangle_{\partial\Omega}.
\end{equation}
Here, $\langle \VF{X}, \VF{\phi} \rangle_{\partial\Omega} = \int_{\partial\Omega} \VF{X}\cdot\VF{\phi} \dif \sigma$
denotes the boundary integral. 
We can now express the boundary traces as volume integrals.
We say that $\VF{E}$ has tangential trace $\VF{K_E}$ along $\Sigma^{E}$,
if for all smooth vector fields $\VF{W}$ that vanish in a neighborhood of $\Sigma^{M}$ we have 
\begin{align*} 
    \langle \nabla\times\VF{H}, \VF{W} \rangle &= \langle \VF{H}, \nabla\times\VF{W} \rangle + \langle {\VF{K_E}},\VF{W} \rangle_{\partial\Omega} 
    .
\end{align*} 
Analogously, 
we say that $\VF{H}$ has tangential trace $\VF{K_M}$ along $\Sigma^{M}$,
if for all smooth vector fields $\VF{V}$ that vanish in a neighborhood of $\Sigma^{E}$ we have 
\begin{align*} 
    \langle \nabla\times\VF{E}, \VF{V} \rangle &= \langle \VF{E}, \nabla\times\VF{V} \rangle + \langle {\VF{K_M}},\VF{V} \rangle_{\partial\Omega} 
    .
\end{align*} 
We write $H(\curl;\Omega,\Sigma^{E})$ and $H(\curl;\Omega,\Sigma^{M})$
for the closed subspaces of $H(\curl;\Omega)$ and $H(\curl;\Omega)$
that have vanishing tangential traces along $\Sigma^{E}$ and $\Sigma^{M}$, respectively.
\\

To get a \emph{variational form} (also called a \emph{weak form}) for our finite element implementation,
we multiply equations~\eqref{eq:ampereEH}--\eqref{eq:faradayEH} by test vector fields~\cite{LangtangenLogg2017} 
and integrate the equations over $\Omega$.
Therefore, we look for $\VF{E} \in H(\curl;\Omega)$ and $\VF{H} \in H(\curl;\Omega)$ such that 
\begin{align*} 
 &
 \langle \iunit \omega \epsilon \VF{E} + \iunit \omega \xi \VF{H} - \nabla\times\VF{H}, \VF{V} \rangle 
= -\langle \VF{J}_{E} + \iunit \omega \VF{P}, \VF{V} \rangle
 ,
 \\
 &
 \langle \iunit \omega \zeta \VF{E} + \iunit \omega \mu \VF{H} + \nabla\times\VF{E}, \VF{W} \rangle
= -\langle \VF{J}_{M} + \iunit \omega \VF{M}, \VF{W} \rangle
 ,
\end{align*} 
holds for all $\VF{V} \in H(\curl;\Omega)$ and $\VF{W} \in H(\curl;\Omega)$,
and such that we have the boundary conditions 
\begin{align*}
	\normal\times\VF{E}&=\VF{K_M} \textrm{ along } \Sigma^{E},
	\\
	\normal\times\VF{H}&=\VF{K_E} \textrm{ along } \Sigma^{M}.
\end{align*}
In particular, we can reduce this to an equivalent formulation where the unknown variables satisfy homogeneous tangential boundary conditions along their respective boundary parts.
Suppose that $\VF{E}^{\Sigma} \in H(\curl;\Omega)$ and $\VF{H}^{\Sigma} \in H(\curl;\Omega)$ 
satisfy 
\begin{align*}
	\normal\times\VF{E}^{\Sigma} &= \VF{K_M} \textrm{ along } \Sigma^{E}
	,
	\\
	\normal\times\VF{H}^{\Sigma} &= \VF{K_E} \textrm{ along } \Sigma^{M}
	.
\end{align*}
We write 
\begin{align*}
 \VF{S}_{E}
 &=
 \VF{J}_{E} + \iunit \omega \VF{P} 
 +
 \iunit \omega \epsilon \VF{E}^{\Sigma} + \iunit \omega \xi \VF{H}^{\Sigma} - \nabla\times\VF{H}^{\Sigma}
 ,
 \\
 \VF{S}_{M}
 &=
 \VF{J}_{M} + \iunit \omega \VF{M} 
 +
 \iunit \omega \zeta \VF{E}^{\Sigma} + \iunit \omega \mu \VF{H}^{\Sigma} + \nabla\times\VF{E}^{\Sigma}
 .
\end{align*}
Then we search for
$$
\VF{E} \in H(\curl;\Omega,\Sigma^{E}), \quad \VF{H} \in H(\curl;\Omega,\Sigma^{M})
$$
such that 
\begin{align*} 
 &
 \langle \iunit \omega \epsilon \VF{E} + \iunit \omega \xi \VF{H} - \nabla\times\VF{H}, \VF{V} \rangle 
 =
 -\langle \VF{S}_{E}, \VF{V} \rangle
 ,
 \\
 &
 \langle \iunit \omega \zeta \VF{E} + \iunit \omega \mu \VF{H} + \nabla\times\VF{E}, \VF{W} \rangle
 =
 -\langle \VF{S}_{M}, \VF{W} \rangle
\end{align*} 
for all $\VF{V} \in H(\curl;\Omega,\Sigma^{E})$ and $\VF{W} \in H(\curl;\Omega,\Sigma^{M})$.
\\

Let $\bbX_{h}^{E} \in H(\curl;\Omega,\Sigma^{E})$ and $\bbX_{h}^{M} \in H(\curl;\Omega,\Sigma^{M})$.
The corresponding Galerkin problem asks for vector fields 
$\VF{E}_{h} \in \bbX_{h}^{E}$ and $\VF{H}_{h} \in \bbX_{h}^{M}$
such that 
\begin{align*} 
 &
 \langle \iunit \omega \epsilon \VF{E}_{h} + \iunit \omega \xi \VF{H}_{h} - \nabla\times\VF{H}_{h}, \VF{W}_{h} \rangle 
 =
 -\langle \VF{S}_{E}, \VF{W}_{h} \rangle
 ,
 \\
 &
 \langle \iunit \omega \zeta \VF{E}_{h} + \iunit \omega \mu \VF{H}_{h} + \nabla\times\VF{E}_{h}, \VF{V}_{h} \rangle
 =
 -\langle \VF{S}_{M}, \VF{W}_{h} \rangle
 .
\end{align*} 
holds for all $\VF{V}_{h} \in \bbX_{h}^{E}$ and $\VF{W}_{h} \in \bbX_{h}^{M}$.

Our Galerkin method is a finite element method. 
There are various curl-conforming finite element methods for different types of triangulations. 
We will consider N\'ed\'elec edge elements of the first kind with respect to a tetrahedral mesh of the domain\cite{angermann2019semi,monk1992analysis,monk2019finite,chen2000finite,zaglmayr2006high}
as choices of $\bbX_{h}^{E}$ and $\bbX_{h}^{M}$, as these are widely documented in the literature.

\section{\label{sec:numerics} Numerical results}
In this section we discuss the results of numerical computations 
using our finite element method. 
For our calculations, we have used the \emph{Python} package \emph{Netgen/NGSolve} (v6.2.2008) \cite{schoberl2014c++,zaglmayr2006high} 
\\

Our test scenarios have the following form. 
We let $\Omega = [0,1]^{3}$ be the unit cube. 
We consider the generalized Maxwell system with zero magnetization and polarization,
$\VF{P} = \VF{M} = (0,0,0)$, 
frequency $\omega = 1$, 
and different material scalar coefficients. 
We study how the errors of numerically computed electric and magnetic fields behave for four simple scenarios:
\begin{enumerate}
 \item $(\epsilon,\mu) = 1$ and $(\xi,\zeta) = 0$,
 \item $(\epsilon,\mu) = 0$ and $(\xi,\zeta) = 1$, 
 \item $(\epsilon,\mu) = 1$ and $(\xi,\zeta) = 1$,
 \item $(\epsilon,\mu) = 1$ and $(\xi,\zeta) = 0.01$.
\end{enumerate}
The last case $(\epsilon,\mu) = 1$ and $(\xi,\zeta) = 0.01$ could be considered physically interesting because
it involves the standard vacuum material tensors together with a minor non-standard perturbation (as reasonably expected in nature).

We choose the solution fields $\VF{E}$ and $\VF{H}$ to be 
\begin{align*} \begin{split}
        \VF{E} &= 
        \begin{bmatrix}
            \sin( 6\pi x ) \sin( 10\pi y ) \sin( 14\pi z )
            \\
            0
            \\
            0 
        \end{bmatrix},
        \\
        \VF{H} &= 
        \begin{bmatrix}
            \sin( 8\pi x ) \sin( 12 \pi y ) \sin( 2\pi z )
            \\
            0
            \\
            0 
        \end{bmatrix}.
    \end{split}
\end{align*}
Here, $(x,y,z)$ are the spatial coordinates.
Accordingly, one obtains the right-hand sides $\VF{J_E}$ and $\VF{J_M}$ using \eqref{eq:ampereEH}-\eqref{eq:faradayEH}:
\begin{align*} \begin{split}
        \VF{J}_\VF{E} &= 
        \begin{bmatrix}
            - \iunit \omega ( \epsilon \Gamma_1 + \xi \Gamma_2 )
            \\
            2 \pi \sin(8\pi x) \sin(12\pi y) \sin(2\pi z)
            \\
            -12 \pi \sin(8\pi x) \sin( 12\pi y) \sin(2\pi z)
        \end{bmatrix},
        \\
        \VF{J}_\VF{M} &= 
        \begin{bmatrix}
            - \iunit \omega ( \mu \Gamma_2 + \zeta \Gamma_1) 
            \\
            -14 \pi \sin(6\pi x) \sin(10\pi y) \sin(14\pi z) 
            \\
            10 \pi\sin(6\pi x) \sin(10\pi y) \sin(14\pi z)  
        \end{bmatrix},
    \end{split}
\end{align*}
where we abbreviate 
\begin{align*}
    \Gamma_1 &= \sin(6\pi x) \sin(10\pi y) \sin(14\pi z),
    \\
    \Gamma_2 &= \sin(8\pi x) \sin(12\pi y) \sin( 2\pi z).
\end{align*}
Notably, our fields are trivially zero at the boundary of the unit cube.
\\

\begin{figure*} 
    \begin{subfigure}{.5\textwidth}
    \centering
    \includegraphics[width=.8\linewidth]{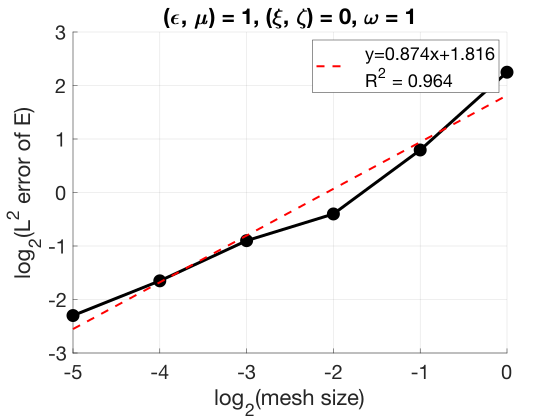}
    \label{fig:errorA:3}
    \end{subfigure}
    \begin{subfigure}{.5\textwidth}
    \centering
    \includegraphics[width=.8\linewidth]{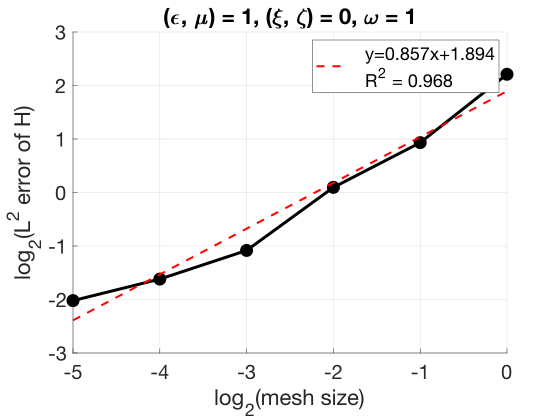}\label{fig:errorA:4}
    \end{subfigure}
    \caption{Behavior of the $L^2$ error
    of numerically computed electric field $\VF{E}$ and magnetic field $\VF{H}$ with decreasing mesh size
    for our standard example.
    Here, we have used $(\epsilon,\mu) = 1$ and $(\xi,\zeta) = 0$.}
    \label{fig:errorA}
\end{figure*}

\begin{figure*} 
    \begin{subfigure}{.5\textwidth}
        \centering
        \includegraphics[width=.8\linewidth]{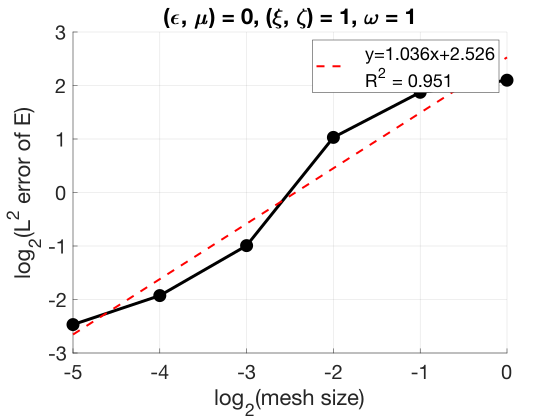}
        \label{fig:errorB:3}
    \end{subfigure}
    \begin{subfigure}{.5\textwidth}
        \centering
        \includegraphics[width=.8\linewidth]{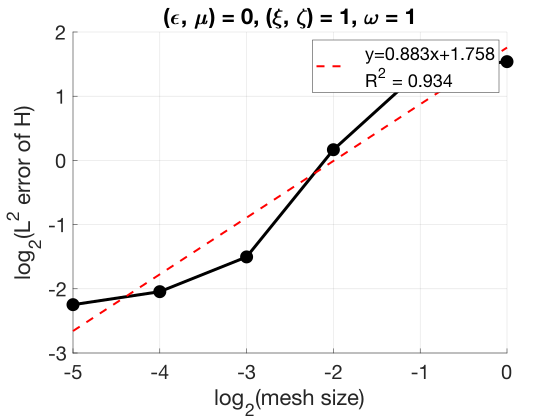}\label{fig:errorB:4}
    \end{subfigure}
    \caption{Behavior of the $L^2$ error
    of numerically computed electric field $\VF{E}$ and magnetic field $\VF{H}$ with decreasing mesh size
    for our standard example.
    Here, we have used $(\epsilon,\mu) = 0$ and $(\xi,\zeta) = 1$.}
    \label{fig:errorB}
\end{figure*}

\begin{figure*} 
    \begin{subfigure}{.5\textwidth}
        \centering
        \includegraphics[width=.8\linewidth]{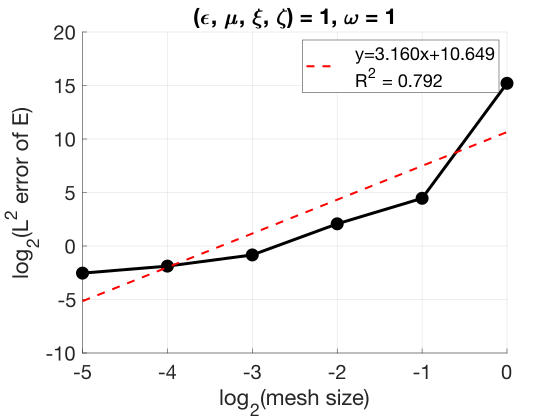}
        \label{fig:errorC:3}
    \end{subfigure}
    \begin{subfigure}{.5\textwidth}
        \centering
        \includegraphics[width=.8\linewidth]{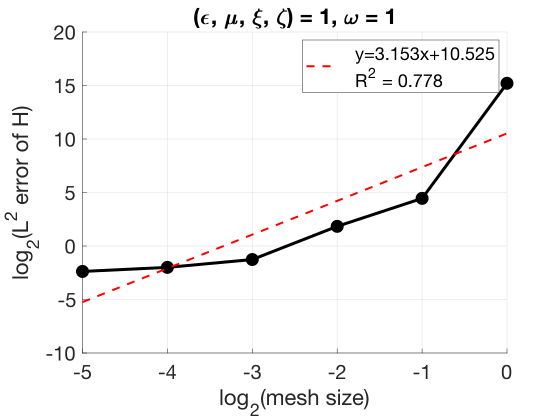}\label{fig:errorC:4}
    \end{subfigure}
    \caption{Behavior of the $L^2$ error
    of numerically computed electric field $\VF{E}$ and magnetic field $\VF{H}$ with decreasing mesh size
    for our standard example.
    We used  $(\epsilon,\mu) = 1$ and $(\xi,\zeta) = 1$.}
    \label{fig:errorC}
\end{figure*}

\begin{figure*} 
    \begin{subfigure}{.5\textwidth}
        \centering
        \includegraphics[width=.8\linewidth]{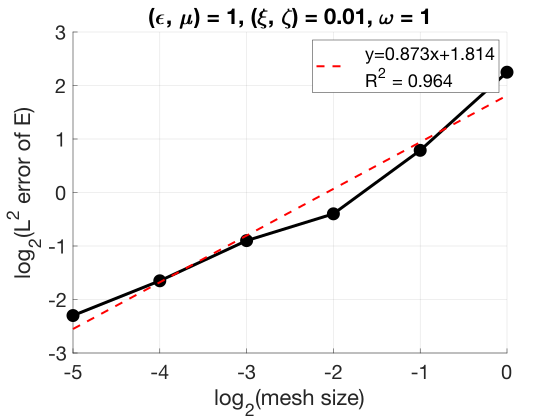}
        \label{fig:errorD:3}
    \end{subfigure}
    \begin{subfigure}{.5\textwidth}
        \centering
        \includegraphics[width=.8\linewidth]{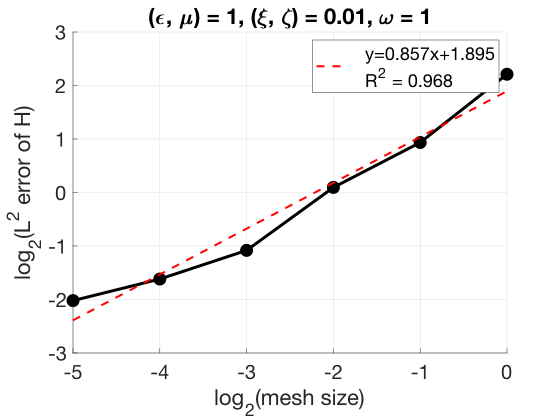}\label{fig:errorD:4}
    \end{subfigure}
    \caption{Behavior of the $L^2$ error
    of numerically computed electric field $\VF{E}$ and magnetic field $\VF{H}$ with decreasing mesh size
    for our standard example.
    We used  $(\epsilon,\mu) = 1$ and $(\xi,\zeta) = 0.01$.}
    \label{fig:errorD}
\end{figure*}

Starting with a simple initial triangulation, we use these choices of $\VF{P}$, $\VF{M}$, $\VF{J_E}$ and $\VF{J_M}$ in the weak formulation and approximate $\VF{E}$ and $\VF{H}$ numerically. 
We compute the finite element approximations up to machine precision for sequences of meshes obtained by uniform refinement. 
The development of the error terms is shown in the $\log\log$ plots in Figures~\ref{fig:errorA}-\ref{fig:errorD}.
The x-axis is the binary logarithm of the mesh size, and the y-axes is the error norm of the two fields $\VF{E}$ and $\VF{H}$.
An upward-slope indicates convergence because the error decreases as the mesh size decreases.
The slope of the linear trendline in each plot approximates the order of convergence for small mesh sizes. 
\\

Figures~\ref{fig:errorA}-\ref{fig:errorD} show that all errors converge. 
In Figure~\ref{fig:errorC}, the error appears to stall
with decreasing mesh size. This is very likely because the 
transformation matrix from \eqref{eq:genconst0a}-\eqref{eq:genconst0b} --
$$
\begin{bmatrix}
\epsilon & \xi \\
\zeta & \mu 
\end{bmatrix}
$$
-- is not invertible when $(\epsilon,\mu,\xi,\zeta) = 1$.
For completeness, we present the raw data used for 
Figures~\ref{fig:errorA}-\ref{fig:errorD} in 
Tables~\ref{tab:case1}-\ref{tab:case4}.

\begin{table}[]
\centering
    \begin{tabular}{|c|c|c|}
        \hline
        $1/h$ & $L^2(E)$ & $L^2(H)$ \\ \hline
        1     & 4.75E+00 & 4.63E+00 \\ \hline
        2     & 1.73E+00 & 1.91E+00 \\ \hline
        3     & 7.57E-01 & 1.07E+00 \\ \hline
        4     & 5.36E-01 & 4.72E-01 \\ \hline
        5     & 3.18E-01 & 3.26E-01 \\ \hline
        6     & 2.03E-01 & 2.46E-01 \\ \hline
    \end{tabular}
    \caption{$L^2$ errors in the fields $E$ and $H$ for the case $(\epsilon,\mu) = 1$ and $(\xi,\zeta) = 0$.}
    \label{tab:case1}
    \end{table}

\begin{table}[]
\centering
    \begin{tabular}{|c|c|c|}
        \hline
        $1/h$ & $L^2(E)$ & $L^2(H)$ \\ \hline
        1     & 4.28E+00 & 2.90E+00 \\ \hline
        2     & 3.66E+00 & 2.62E+00 \\ \hline
        3     & 2.04E+00 & 1.12E+00 \\ \hline
        4     & 5.02E-01 & 3.53E-01 \\ \hline
        5     & 2.63E-01 & 2.42E-01 \\ \hline
        6     & 1.81E-01 & 2.11E-01 \\ \hline
    \end{tabular}
    \caption{$L^2$ errors in the fields $E$ and $H$ for the case $(\epsilon,\mu) = 0$ and $(\xi,\zeta) = 1$.}
    \label{tab:case2}
    \end{table}

\begin{table}[]
\centering
    \begin{tabular}{|c|c|c|}
        \hline
        $1/h$ & $L^2(E)$ & $L^2(H)$ \\ \hline
        1     & 3.77E+04 & 3.77E+04 \\ \hline
        2     & 2.20E+01 & 2.19E+01 \\ \hline
        3     & 4.21E+00 & 3.58E+00 \\ \hline
        4     & 5.61E-01 & 4.16E-01 \\ \hline
        5     & 2.72E-01 & 2.50E-01 \\ \hline
        6     & 1.73E-01 & 1.92E-01 \\ \hline
    \end{tabular}
    \caption{$L^2$ errors in the fields $E$ and $H$ for the case $(\epsilon,\mu) = 1$ and $(\xi,\zeta) = 1$.}
    \label{tab:case3}
\end{table}

\begin{table}[]
    \centering
    \begin{tabular}{|c|c|c|}
        \hline
        $1/h$ & $L^2(E)$ & $L^2(H)$ \\ \hline
        1     & 4.75E+00 & 4.63E+00 \\ \hline
        2     & 1.73E+00 & 1.91E+00 \\ \hline
        3     & 7.59E-01 & 1.07E+00 \\ \hline
        4     & 5.36E-01 & 4.73E-01 \\ \hline
        5     & 3.19E-01 & 3.26E-01 \\ \hline
        6     & 2.03E-01 & 2.46E-01 \\ \hline
    \end{tabular}
    \caption{$L^2$ errors in the fields $E$ and $H$ for the case $(\epsilon,\mu) = 1$ and $(\xi,\zeta) = 0.01$.}
    \label{tab:case4}
\end{table}

\section*{Acknowledgments}
The authors thank Peter Monk for helpful correspondence.
This material is based on work supported by the National Science Foundation under Grant No.\ DMS-1439786 while the second author was in residence at the Institute for Computational and Experimental Research in Mathematics in Providence, RI, during the ``Advances in Computational Relativity'' program.

\end{document}